\documentclass[final]{aipproc}

\layoutstyle{6x9}

\begin{document}

\title{What did we learn and what will we learn from hydrodynamics at RHIC?}

\author{Peter F. Kolb}{
  address={Department of Physics and Astronomy, 
                    SUNY Stony Brook, Stony Brook, NY 11974, USA}
}

\begin{abstract}


A brief overview of the current status of hydrodynamic concepts 
applied to ultra relativistic heavy-ion collisions is presented. 
Special emphasis is placed on future prospects for extracting the 
 thermodynamic properties and the bulk features of the created medium from
experimental observations.

\end{abstract}

\maketitle


An extensive review of the application of hydrodynamics 
to describe the expansion stage of ultra relativistic heavy-ion collisions
was recently given elsewhere \cite{KH03}.
Therefore I will be brief in reporting the highlights of past studies, but
emphasize more what can be expected from the future application 
of hydrodynamics. 
For the basic concepts of hydrodynamics,
its formalism and a complete  list of references
see also  \cite{KH03}.

\section{Anisotropic flow and hydrodynamics at RHIC}

One of the first and still most striking observations at RHIC
is a strong event anisotropy in non-central collisions \cite{STARfirstv2},
which is generated through the elliptically deformed overlap region
of the colliding nuclei, resulting in an eccentric distribution of matter
and anisotropic pressure gradients in the early stages of the expansion
\cite{Ollitrault92}.
From a microscopic point of view this strong collective anisotropy 
is best described under the assumption of extremely strong 
rescattering \cite{MG02}, strong enough in fact to reach the limit of 
continuum dynamics, whose predictions \cite{KSH00} were in quantitative 
agreement with first observations \cite{STARfirstv2}. 
To achieve such a strong conversion of anisotropies from coordinate 
to momentum space, rescattering has to be strong at very early times 
and local thermalization has to occur while the geometric deformation 
of the source is still large \cite{KSH00}.
More detailed subsequent hydrodynamic studies of non-central
collisions gave  additional predictions on characteristic features such as the mass 
dependence of elliptic flow \cite{Kolbetalv2} and  investigated the influence
of characteristics of the underlying equation of state \cite{Kolbetalv2,Teaneyetalv2}. 
Results of such studies are given in Figure 1, 
where the left panel shows the momentum dependence
of elliptic flow $v_2(p_T) = \langle \cos 2 \varphi \rangle$, 
the second Fourier coefficient of the azimuthal dependence of the particle spectra
$\frac{dN}{p_T dp_T dy d\varphi}$, as a function of transverse momentum for pions and 
protons. 
Experimental data from minimum bias collisions \cite{STARidv2} 
are compared to hydrodynamic results \cite{Kolbetalv2,HK02},  
once applying an equation of state including a phase 
transition to a plasma stage (solid lines), 
once using a soft resonance gas equation even at the earliest and hottest
stages of the collision (dashed lines). 
The right panel of the figure shows the average elliptic flow of all 
charged particles as a function of centrality. Included are results of a 
hydrodynamic calculation 
assuming different latent heats $\Delta e$ of the transition region 
(0.4, 0.8 and 1.6 GeV/fm$^3$) \cite{Teaneyetalv2}. 
Clearly these results indicate that in order to describe the absolute
magnitude, the mass splitting and the centrality dependence of elliptic flow,
rapid thermalization with a strong push from a phase with a sufficiently 
hard equation of state (like the QCD plasma) and a fairly soft transition region 
(of width $\Delta e \sim 1$~GeV/fm$^3$) back to hadronic matter is required.

%
 \begin{figure}[h,t,b,p]
 \begin{minipage}{8cm}
 \includegraphics[width=7.cm]{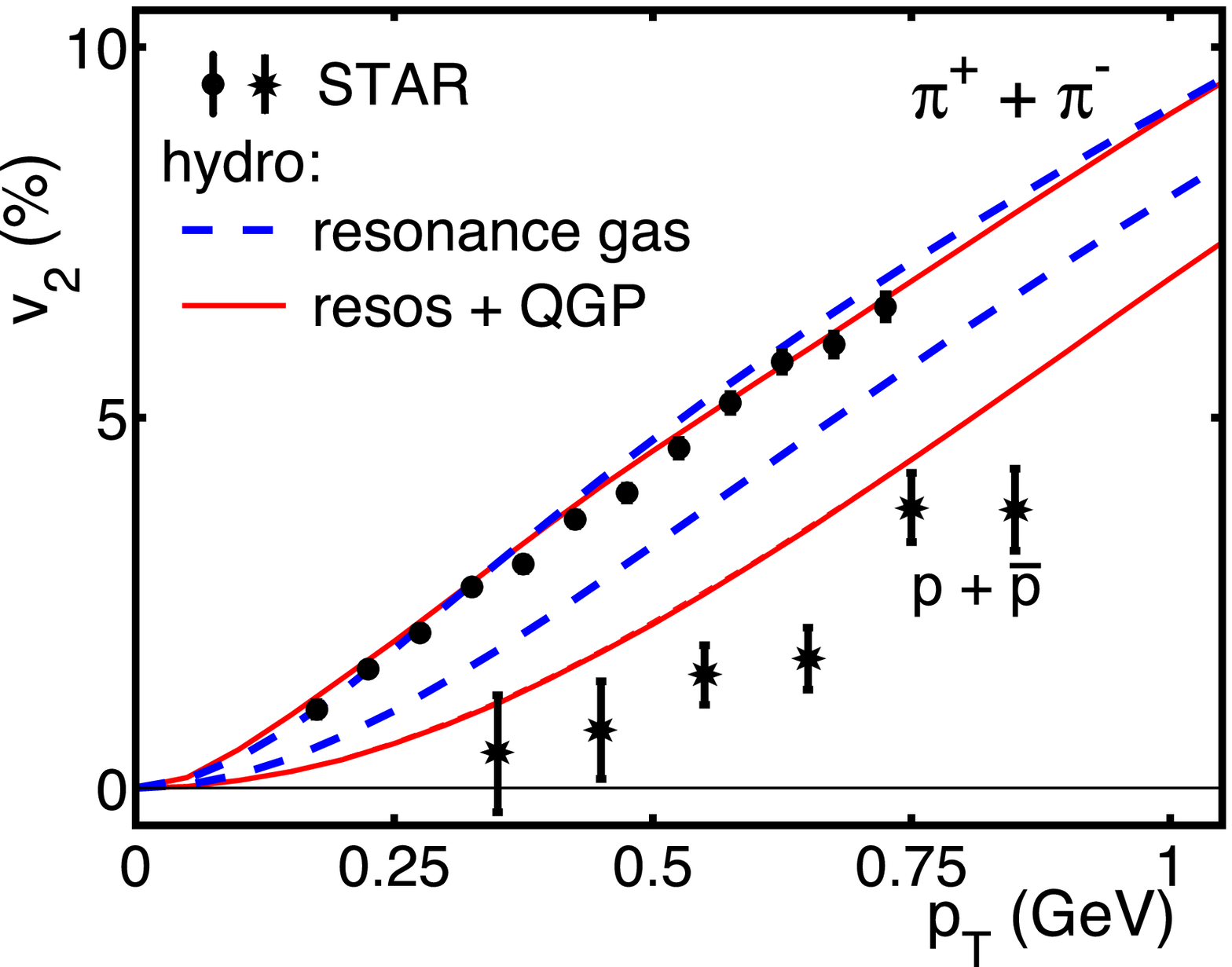}
 \end{minipage}
 \begin{minipage}{6cm}
 \includegraphics[width=5.5cm]{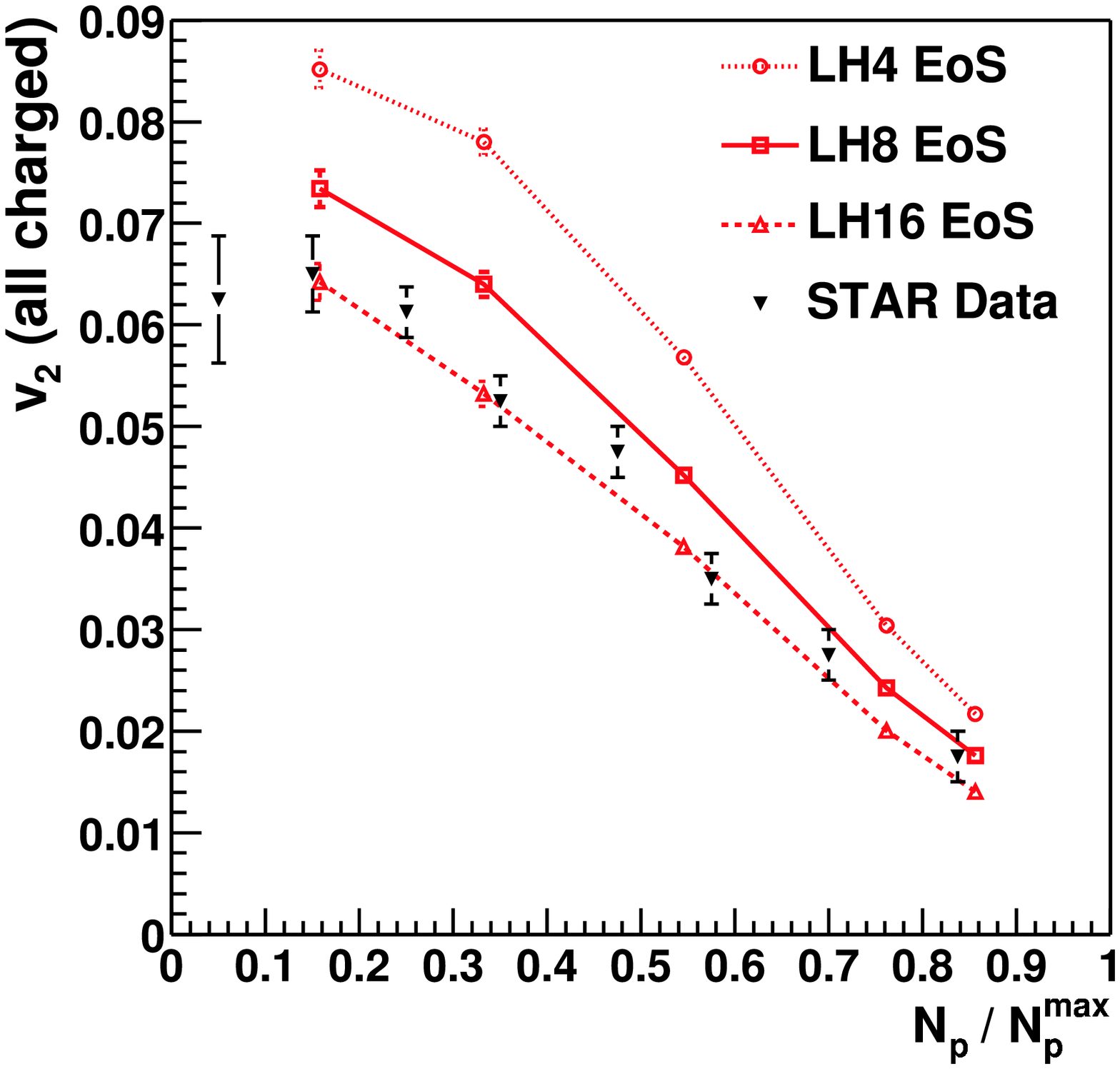}
 \end{minipage}\\
 \caption{
        Experimental data of elliptic flow from Au+Au collisions at 130 AGeV     	\cite{STARidv2,STARfirstv2}
        in comparison to results from hydrodynamic calculations.
        Left: Elliptic flow as function of transverse momentum of pions and protons in 
                 minimum bias  from a purely hydrodynamic calculation \cite{Kolbetalv2,HK02},
                  applying an
                 equation of state with (solid) and without a phase transition (dashed).      
        Right: centrality dependence of elliptic flow of a hydrodynamic calculation which
                  treats the late hadronic stage within a microscopic picture. 
                  Results using different widths of the transition region 
                  (0.4, 0.8 and 1.6 GeV/fm$^3$) are shown \cite{Teaneyetalv2}.\vspace*{-.5cm}}
 \label{fig:polarplots}
 \end{figure}


\section{The road ahead}

With the prerequisites of a thermalized, adiabatically evolving source given,
there is a wealth of topics that we can address within the hydrodynamic framework.

\begin{figure}[h,t,b,p]
 \hspace*{-.500cm}
 \begin{minipage}{6cm}
 \includegraphics[width=6cm]{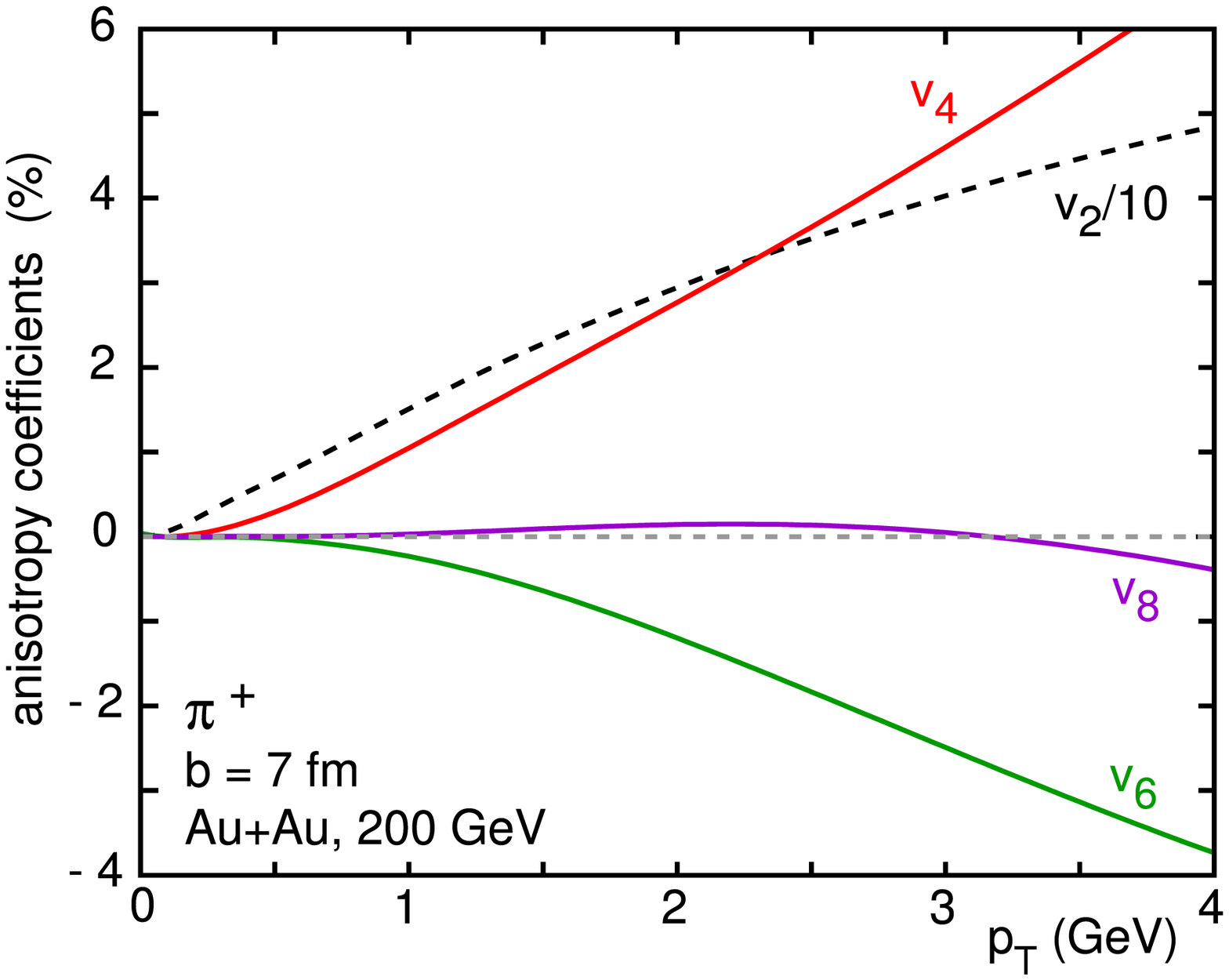}
 \end{minipage}
 \hspace*{.5cm}
 \begin{minipage}{7cm}
 \includegraphics[width=7.5cm]{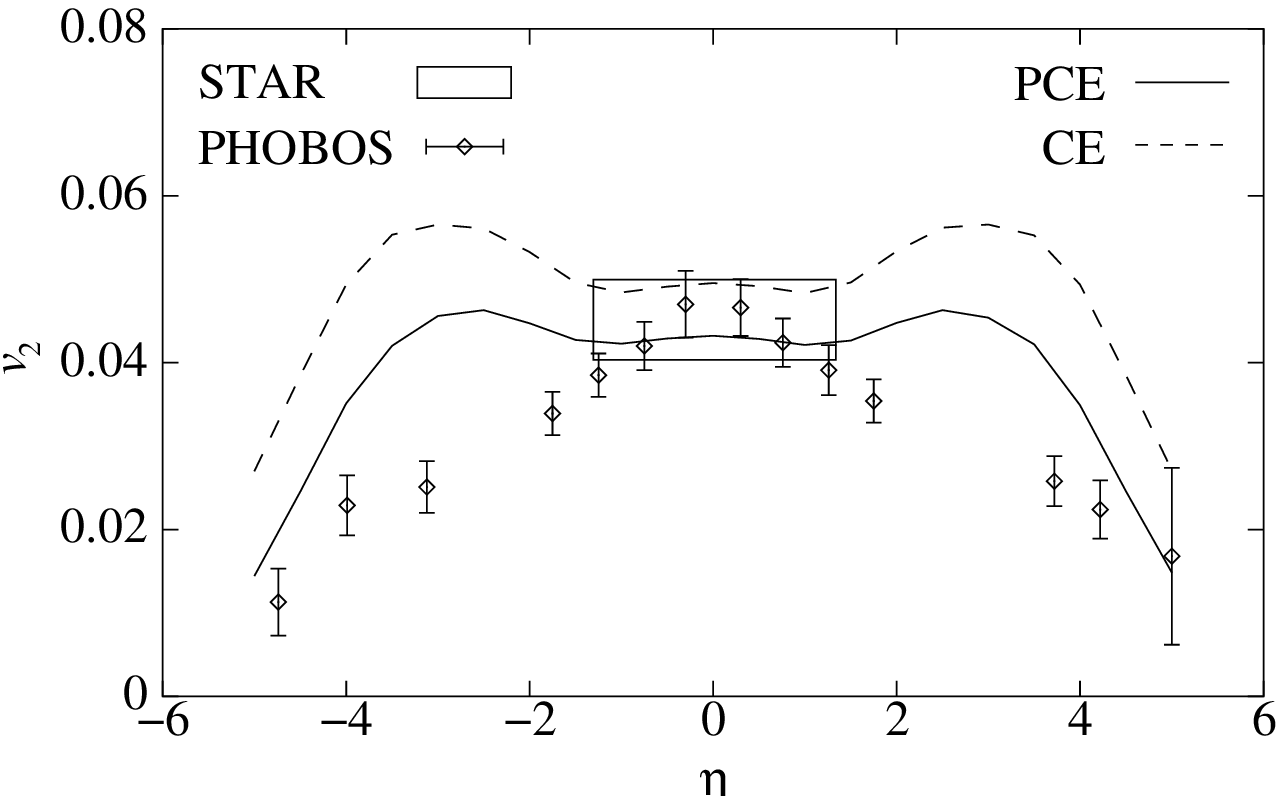}
 \end{minipage}\\
 \caption{
        Left: Hydrodynamic expectations for higher flow anisotropies, 
               calculated for pion spectra resulting from Au+Au collisions with 
               $b=7$~fm 
                  \cite{Kolb03}.
        Right: rapidity dependence of elliptic flow from a hydrodynamic model \cite{Hirano02}
                   (with different degrees of chemical equilibration in the resonance stage),
                   compared to experimental data from STAR and PHOBOS at 130 GeV
                   \cite{STARfirstv2,PHOBOSv2}. \vspace*{-4mm}}
 \label{fig:Fig2}
 \end{figure}

{\bf More quantitative extraction of the properties of the equation of state:}
Hydrodynamics is {\em the tool} to study how properties of the equation of state
influence the dynamics of the system and final state observables, as the
equation of state, which relates the local energy density to the pressure,
explicitly enters the formalism in terms of the forces that drive the
system apart.
With the large variety of flow observables becoming currently available, 
we will be  able to delimit parameters of the calculation, and particularly 
put stronger constraints on features of the nuclear equation of state by 
comparing experimental data with theoretical calculations. 
Helpful in this context are also finer details in the particle emission pattern, 
such as anisotropic components $v_n = \langle \cos n \varphi \rangle$
beyond the elliptic deformation,
which may achieve significantly large values at intermediate to high transverse
momenta \cite{Kolb03}. The left panel of Fig. \ref{fig:Fig2} shows the momentum
dependence of the Fourier coefficients up to order 8 as expected from a 
hydrodynamic calculation to describe Au+Au collisions at $\sqrt{s_{\rm NN} }=200$~GeV.

%

%
{\bf Collectivity of which particles at what stages?}
The most abundant hadrons at RHIC -- pions, kaons and protons -- 
share signs of a strong common transverse expansion. 
Newest results on particle spectra of  strange and multi-strange baryons are 
yet inconclusive to which extent these species follow the collective dynamics. 
While the traditional  'blast wave fit' tends to assign them a higher 
temperature of origin \cite{strangenessflow}, they appear 
to follow the common mass systematics of the full hydrodynamic calculation
down to freeze-out temperatures of 100 MeV \cite{KH03}.
Signals attributed to decaying charmed mesons \cite{PHENIXel} 
do not yet prove or disprove whether even heavy flavors participate in 
the strong collectivity \cite{BKGN03}.
Whereas the question of flow of strangeness in the hadronic phase is still 
debatable, there is clear evidence from flow anisotropies of strange particles
\cite{strangev2exp}  that at hadronization all quark flavors share
the same radial as well as anisotropic flow \cite{Kolbetalv2,v2coalescence}, 
providing a clear sign of early thermalization and a common 
collective expansion of the fireball in the partonic stage. 


{\bf Equation of state at larger baryon densities:}
Flow observables at forward/backward rapidities open the possibility to
study the nuclear equation of state at larger baryon densities than found 
at central rapidity, and thus allow to scan a larger portion of the 
$(T,\mu)$-plane of our nuclear equation of state, eventually bringing us 
into the realm of the expected tri-critical point \cite{FK02}.
This issue is however complicated through the breakdown 
of thermalization when moving toward the fragmentation region 
\cite{Hirano02}. 
Before one can make quantitative statements about the equation of state
in this realm, the exceedingly complicated interplay of initial conditions, 
viscous effects and the equation of state have to be well understood.
Results from a hydrodynamic calculation that extends over a large rapidity 
window are shown in the right panel of Fig. \ref{fig:Fig2} \cite{Hirano02}. 
Deviations from ideal hydrodynamic behavior occur at rapidities 
$| \eta | > 1$ from where on the mentioned effects have to be taken into account.
%


{\bf Viscosity effects:}
Clearly the approach of ideal hydrodynamics  works only under the stringent
conditions of local thermalization followed by a 'gradual' adiabatic expansion. 
Deviations from this behavior are expected in the most peripheral
collisions, when approaching the fragmentation region, 
in the late stages of the reaction and for the few particles emitted with large 
transverse momenta. 
Under these conditions, viscosity effects need to be considered \cite{Teaney03}. 
Although the treatment of those effects within a full dynamical calculation is
a very difficult task \cite{Muronga02}, it will eventually enrich our understanding of the 
transport properties of the quark gluon plasma and the hadron gas, and maybe resolve 
the persisting HBT-puzzle at RHIC \cite{HBTpuzzle}.


{\bf Background medium for hard probes:}
Hydrodynamics has proved to be a great tool to study the properties of the bulk of
the expanding matter. 
Even hard probes, although they do not follow the collective dynamics of the bulk,
depend on the dynamical evolution of the fireball. 
The characteristics of energy-loss and jet-quenching should thus be folded into the
hydrodynamic expansion, to get a reliable description of the net energy loss which 
the hard probes experience during the fireball evolution \cite{hydroandjet}.

\section{Summary}

To address the  thermodynamic properties of the medium created at RHIC, 
it is essential that the system rapidly achieves local thermal equilibrium,
which appears to happen within the first 1 fm/$c$ after impact.
The most natural language to study the nuclear equation of state, its influence 
on the dynamics of the system and the resulting observables is hydrodynamics.
With the steady output of flow observables from RHIC we can constrain
the parameters of the equation of state which is responsible for the observed 
strong collective expansion, but also address questions of viscosity and other
non-ideal effects.

{\bf Acknowledgments:}
This work was supported in parts by the U.S. Department of Energy under
Grant No. DE-FG02-88ER40388. Support from the Alexander von Humboldt 
Foundation in terms of a Feodor Lynen Fellowship is greatly appreciated.
\\[-.7cm]

\bibliographystyle{aipprocl} 


\IfFileExists{\jobname.bbl}{}
 {\typeout{}
  \typeout{******************************************}
  \typeout{** Please run "bibtex \jobname" to optain}
  \typeout{** the bibliography and then re-run LaTeX}
  \typeout{** twice to fix the references!}
  \typeout{******************************************}
  \typeout{}
 }

\end{document}